\documentclass[12pt]{article}
\usepackage{graphicx}
\usepackage{amssymb,amsfonts,amsmath}
\usepackage{natbib}

\begin{document}

\title{Power laws and fragility in flow networks}

\author{Jesse Shore$^{a\dagger*}$, Catherine J. Chu$^{bc}$ and Matt T. Bianchi$^{bc}$\\
\small$^{a}$Boston University\\
\small$^{b}$Massachusetts General Hospital\\
\small$^{c}$Harvard Medical School\\
\small$^{\dagger}$jccs@bu.edu}

\maketitle
\begin{abstract}What makes economic and ecological networks so unlike other highly skewed networks in their tendency toward turbulence and collapse? Here, we explore the consequences of a defining feature of these networks: their nodes are tied together by flow. We show that flow networks tend to the power law degree distribution (PLDD) due to a self-reinforcing process involving position within the global network structure, and thus present the first random graph model for PLDDs that does not depend on a rich-get-richer function of nodal degree.  We also show that in contrast to non-flow networks, PLDD flow networks are dramatically more vulnerable to catastrophic failure than non-PLDD flow networks, a finding with potential explanatory power in our age of resource- and financial-interdependence and turbulence. 
\end{abstract}
\section{Introduction}
As the financial crisis of 2008 began to gather force, \cite{may2008complex} observed that economic networks are similar to ecological networks in that both have highly skewed degree distributions, and most worrisome at that time, both are so prone to sudden collapse.   Although there is a rich literature on networks with highly skewed, frequently power-law degree distributions (PLDDs) across many domains of study  \citep{Beyeler2007693,boss2004network,broder2000graph,Dunne01102002,faloutsos1999power,inaoka2004fractal,jeong2000large,newman2005power,paczuski2004heavenly}, existing literature does not provide any adequate models to understand this type of complex phenomenon. A wide range of general graph formation algorithms produce the PLDD  \citep{BarAlb1999,PhysRevE.70.066149,barrat2004weighted,bianconi2001competition,dorogovtsev2004minimal,kleinberg1999web,li2004comprehensive,saramaki2004scale,simon1955class,vazquez2003growing,yook2001weighted}, but each of these models generates some form of  ``rich-get-richer'' attachment: nodes that have many links have an increased probability of adding even more links.  These models trace the connection between degree, local link-addition rules and a global outcome: the overall degree distribution.  

This literature is insufficient to explain the problem posed by \cite{may2008complex} in two ways. First, in social networks defined by flows of resources, one's current degree is not the most natural predictor of future degree.  For example, in economic networks, if one wants to make payments (financial connections) to a large number of other economic actors, the fundamental issue is whether one has the resources with which to do it.  A person could have only one incoming link, but if that link conveys a large amount of money, then that person is in the position to create a large number of outgoing links in the future.  The second and more dramatic difference is that existing literature predicts that PLDD networks should be robust \citep{albert2000error,albert2004structural}, not turbulent and prone to collapse. 

Here, we explore a defining feature of economic and ecological networks: their nodes are tied together by flow.  This means that any disruption of flow has both local and downstream effects that can cascade through the entire network.  We show how throughput, which depends on \emph{position} within the network, rather than current degree, could be a determining factor in the evolution of the overall degree distribution of resource-flow networks.  In so doing, we present what is to our knowledge the first random graph model for PLDDs that does not depend on a rich-get-richer function of nodal degree, and therefore identify a different class of conditions under which the PLDD should be expected due to random processes. Additionally, we show how such PLDD flow networks are prone to catastrophic collapse.

Flow networks are distinct from purely relational networks, in which the links represent a non-fungible connection, such as trust or similarity (Figure 1).  Rather, flow networks are those in which a flowing quantity can pass serially from node to node through the network.  We limit our analysis to the class of flow networks for which flow can be modeled as (approximately) conserved.  That is, although some flow may be lost to transmission or transaction costs, a node's total out-flow cannot be greater than its in-flow and is of the same approximate magnitude.  More specifically, the flow networks in this paper should be distinguished from flow networks in which flow is not conserved, such as information networks \citep{borgatti2005centrality}. Prior models of flow networks have been primarily concerned with characterizing their structure and its implications in cross section \citep{barrat2004architecture,borgatti2005centrality,chung1997spectral, Freeman1991141, holme2003congestion,kleinberg1999authoritative,Newman200539,newman2006modularity}.  Here, we evaluate the dynamics of structure and flow in flow networks as they evolve over time. 

Among social networks, the best examples of flow networks are economic networks.  The movements of any type of fungible asset, be it currency, credit, commodity or equity, could be modeled as this sort of flow network.  This is admittedly a small subset of \emph{social} network phenomena, but a critically important subset, around which many other social networks are organized.    

Our argument proceeds as follows.  In section 2, we illustrate distinguishing features of flow networks.  In section 3, we present an algorithmic model we use to explore the consequences of these distinguishing features.  In section 4, we present our results, and in section 5, we conclude with additional discussion.
  
\section{Modeling Flow Networks}
We begin by considering the defining features of flow networks.  First, in flow networks, the level of flow through each node has consequences for the level of flow through downstream nodes.  As an example, imagine a subnetwork of a food web consisting of strawberries, hares, wolves and foxes, in which the nodes are species, and the links are flows of energy through feeding of one species on another.  Links can have different relative weights and different absolute levels of flow.  For example, although seventy percent of the hares may be eaten by wolves and thirty percent by foxes, the absolute level of energy passing to the wolves and foxes depends on the total number of hares, which in turn depends on the supply of energy from strawberries.  In this toy example, when the strawberries do poorly, so do the wolves, although they are only indirectly connected. In another example, in an inter-city airline transportation network, the number of planes leaving a city does not exceed the number coming in.  Similarly, in banking, a firm's outflow of money is (ultimately) limited by its inflow. 

Second, maintaining relationships generally entails a cost, so if the absolute flow of energy through a relationship dwindles to a negligible level, the relationship may eventually be broken.  If hares become so scarce that it is no longer worth the foxes' effort to hunt for them, they may eventually rely on their other sources of food and sever their link to hares. In the airline transportation network, if the number of passengers between a given city pair is insufficient to cover costs of operation, then a direct route cannot be maintained.  Conversely a ``hub'' city can maintain routes to many cities above the minimum threshold required to cover costs. In general, the number of destination cities that a given airport can serve (its outdegree) is bounded by its own inflow of travelers.  If an airport's role as an intermediary stop for a large number of passengers declined, it may cease to have enough passengers to maintain its lowest traffic routes.  In financial networks, each relationship entails a transaction cost, so very low-value loans are not economical to pursue.

Third, new connections are made over time.  New predator-prey relationships could be formed by necessity, migration, or environmental changes that put new combinations of species in contact.  Exogenous changes to airline demand, as well as operational restructuring can lead to new routes in airline networks which would lead to a redistribution of passenger flow through the network.  Similarly, new relationships are created which result in changes to the structure of financial networks. 
    
 \begin{figure}[h!]
	\centering
		\includegraphics[width=.8\textwidth]{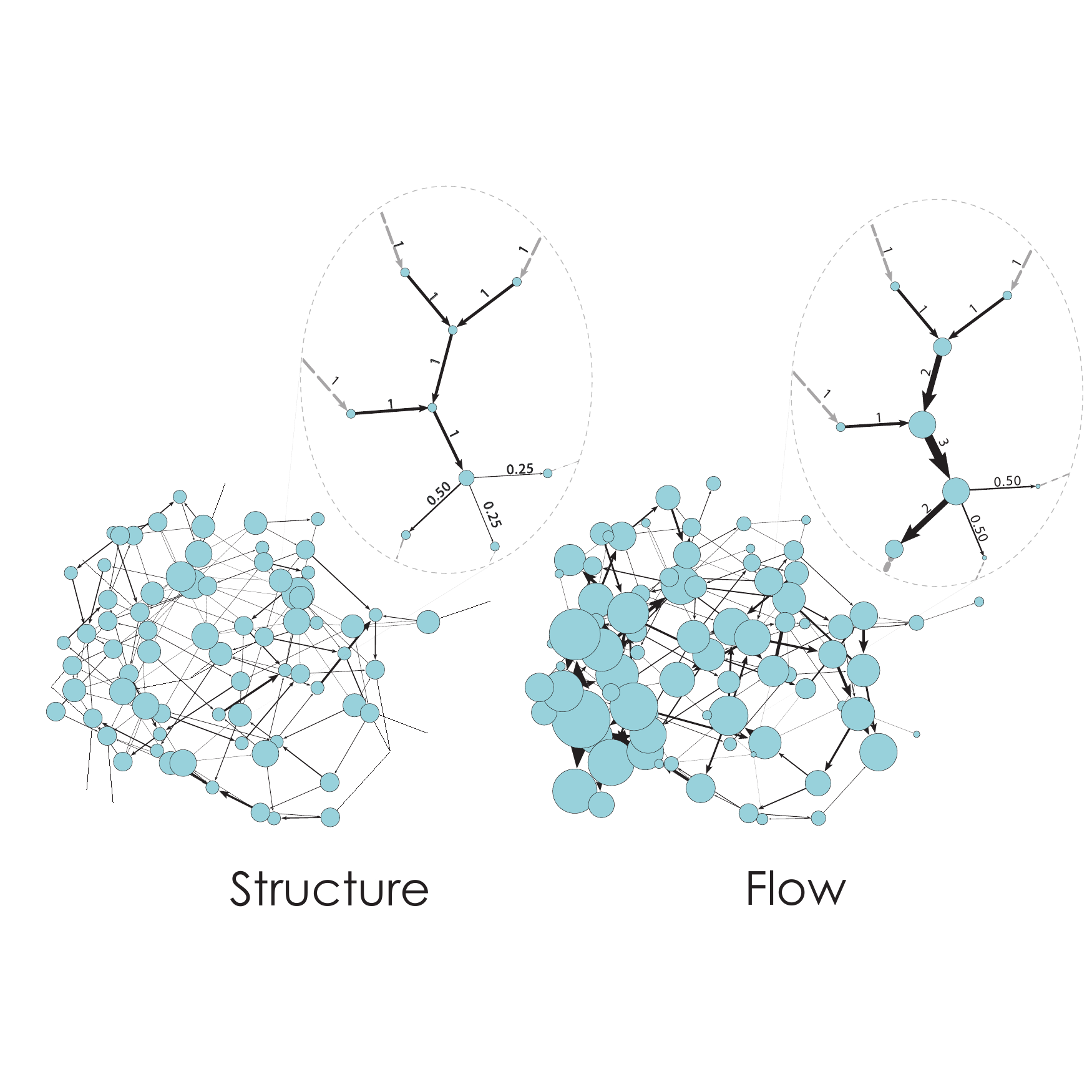}
			\caption{\footnotesize Visualizations of the giant component of an Erd\"{o}s-R\'{e}nyi random graph demonstrating the difference between distributions of degree and flow.  Left: the size of nodes represents their outdegree, and the width of links represents the fraction of the origin node's flow that passes over them.  Right: the sizes of nodes and links represent the absolute amount of flow passing through them, as in a Markov Chain steady state. Insets include labels for weight (left) and the resulting levels of flow (right).  Although all of the links in the top portion of the inset have equal structural weight, their flow varies according to their place in the structure.  In the greater network, flow is concentrated in the neighborhood at the lower-left of the visualization, while structure is more evenly distributed. Visualizations: Pajek \citep{pajek}}
	\label{fig:structure-v-flow}
\end{figure}

Although each real flow network has its own particular features, we identify the following assumptions that provide a general description of this class of networks.   
\begin{enumerate}
	\item Outflow is limited by inflow.  
	\item New relationships continue to form over time.
	\item Very weak links cannot be maintained.  	
\end{enumerate}

For the purposes of the present analysis, we are considering only networks in which the flow is of some fungible resource.  This (along with our assumption 1, above) excludes some flow networks, to which we do not claim our analysis applies.  One such excluded network would be the United States Postal Service.  Personal letters are not interchangeable, and the notion of conservation of flow does not apply.  We also do not analyze multipartite networks and assume that in principle any node could be connected to any other node.

\section{Model}
In order to test these hypotheses and better understand the cumulative effect of the co-evolution between flow and structure, we conducted iterative numerical simulations of dynamic flow networks. 

\subsection{Overview} For each combination of initial parameter conditions($n_{total}$ and $k_{initial}$), we conducted 100 independent simulation runs.  Each run consisted of initialization, iterations and data collection.  Iterations comprised steps for link addition, simulation of flow diffusion, pruning of weak links, and finally testing for fit to a power law. The following paragraphs provide further details for each step.  \\

\subsection{Initialization} Within each simulation, the population of nodes, $n_{total}$, is held constant, but nodes are allowed to become disconnected from and reconnected to the portion of the network with positive throughput.  The average degree of the network is initialized to be $k_{initial}$, but varies as links are added and dissolved.  We represent flow networks with three matrices.  The first, a ``link matrix'' represents the unnormalized weights of each link.  The initial locations of links were determined by sampling uniformly without replacement from the $n^{2}-n$ non-diagonal entries in the matrix.  Weights for these initial links were drawn from the uniform distribution between 0 and 1.

The second matrix represents the local ``structure,'' or the level of flow over the links relative to the total throughput of their origin nodes.  These relative weights are calculated by dividing the absolute weight of an individual link by the sum of all absolute weights of outgoing links from the same origin node, analogous to a row-normalized transition matrix in a Markov chain. The third matrix, the ``flow matrix,'' represents the level of flow over the links, relative to the total flow in the entire network (see below for calculation of this matrix). \\

\subsection{Link addition} We added a small number of links in each iteration. We defined ``small number'' as a random draw from the Poisson distribution with $\lambda=1$, which means that the expected number of links in each iteration was 1, but some iterations had no links added, while others had more than one.  We made the choice to have a varying number of links added at each iteration to ensure that any results were robust to such minor perturbations. 

The origin node for each added link was chosen uniformly randomly from among the nodes with positive throughput.  The destination node was chosen uniformly randomly from among all nodes with which the origin node is not yet connected. The weight of each new link in the structure matrix was defined in the same way as the initial links, by a random draw from the uniform distribution.\\

\subsection{Diffusion of flow} The level of flow is simulated by a numerical approximation to the Markov chain steady state of the transition matrix in each iteration by repeatedly summing the total incoming flow to each node and then dividing this total inflow among its outgoing links according to their normalized weights.  We assume that flow emanates from all nodes, and the process is initialized by equating the flow matrix to the transition matrix and repeated until a stable distribution of flow over the nodes is achieved. ``Stable'' is defined as when the sum of the differences between the throughput in consecutive repetitions is less than one percent of the total throughput in the network. That is, given an $n$ by $n$ transition matrix, $\textbf{T}$, flow matrix, $\textbf{F}$ with entries $\textbf{F}_{i,j,r}$ representing the flow from $i$ to $j$ during repetition $r$, the diffusion sub-steps are initialized by setting $\textbf{F}=\textbf{T}$ and repeated until $$	\displaystyle\sum_{j}^{} \displaystyle\sum_{i}^{} \left|\textbf{F}_{i,j,r+1}-\textbf{F}_{i,j,r}\right|<	\frac{n}{100}$$ or when $r=30$. This level of approximation greatly lessened the computational cost of the simulations.  By comparison, eigenvector decomposition of the transition matrices would have added approximately 4.5 years of processor time for a negligible gain in precision in calculating the steady state over all iterations and all runs. In addition to the computational benefit it generates, we would argue that setting a maximum number of diffusion steps is justifiable in the sense that systems that take exceptionally long to reach their theoretical equilibrium are not likely to reach that equilibrium before the next link in the network changes. We also note that in calculating an equilibrium level of flow, we are assuming no stockpiling or holding of flow within one node occurs at the iteration time scale. \\

\subsection{Link Pruning} Like link addition, in each iteration, we removed a ``small number'' (as above, a new random draw at each iteration) of the weakest links from the network, replacing their entries in all three matrix representations with a 0.  When multiple links were tied for weakest, we selected one uniformly randomly for dissolution.  Rather than use a fixed lower threshold, by simply choosing the number of links to dissolve, we allow the threshold to float, which keeps addition and subtraction of links in approximate balance.

When these dissolutions resulted in the flow over other (downstream) links going to zero, we dissolved these links with non-zero weights but zero flow recursively until none remained. \\   

\subsection{Testing fit of the power law} After every five iterations, we test the distribution of outdegrees for the probability that it was drawn from a power-law statistical distribution with the Kolmogorov-Smirnov test for goodness of fit, comparing it to 100 comparable randomly generated data sets that are known to be drawn from a power-law-distribution \citep{massey1951kolmogorov,clauset2009power}.  The p-value resulting from this test is an approximation of the probability that the distribution of outdegrees fits a power law distribution at least as well as an equivalent random sample (one with the same number of observations $n$, lower cutoff $x_{min}$, and exponent, $\alpha$)  drawn from a true power law distribution.  We based our implementation on code provided by Clauset, et. al., but added the additional restriction that at most $1/3$ of the observed degree distribution be below the lower threshold, $x_{min}$ (see \cite{clauset2009power} for details of the fitting routine), to avoid fitting trivially small portions of the degree distribution to the power law.

It is not easy to definitively establish that an empirical distribution fits a power law rather than another highly skewed distribution.  However, in terms of the empirical phenomena that are characterized by highly skewed degree distributions, what usually demands explanation is not whether the distribution is better described as a power law or the upper tail of a log-normal distribution, it is why some actors or entities are so highly connected, while most have very few connections at all, and what dynamic consequences follow from that pattern of connections.

\subsection{parameter values tested} We examined networks with $n_{total}$ of 128, 256, 512 and 1024 nodes, and  $k_{initial}$ of 2, 4 and 8, for a total of 12 possible combinations of initial network structural conditions. We selected these values for average degree to approximate typical values of empirical networks (In a comparison of 51 social networks, \cite{faust2006comparing} found that the 25th percentile to 75th percentile range for average degree was approximately 2.6 to 6.8 links per node).  For our range of node sizes, we strove to study a relatively wide, but computationally feasible range of values.  We conducted 100 simulation runs for each of these 12 initial conditions.

	\section{Results}

\subsection{The distribution of flow is more skewed than the distribution of degree in flow networks constructed as Erd\"{o}s-R\'{e}nyi random digraphs}  Simulations of flow in static, directed Erd\"{o}s-R\'{e}nyi random graphs with unweighted links showed that the distribution of flow is more skewed than the distribution of outdegree.  The amount of difference varied markedly according to the average degree of the networks. For low average degree networks, the distribution of flow is dramatically more skewed than the distribution of outdegree (Figure 2).  In higher average degree networks, the difference between the distributions of flow and outdegree was slight. 

\paragraph{Interpretation} If flow is conserved (assumption 1), then flow is concentrated by those nodes with fewer (or more unequally weighted) outgoing links than incoming links, and dispersed by nodes with more (or more evenly weighted) outgoing links than incoming links (Figure 1). Flow is therefore directed toward certain parts of a network and diverted from others, producing an unequal distribution of flow for any network other than a regular lattice.  In sparse networks, even when degree does not vary widely, some nodes will have a high flow, because they happen to be downstream from several concentrating nodes, and some will have very low flow because they are downstream from several dispersing nodes.  

\begin{figure}[h!]
	\centering
		\includegraphics[width=0.9\textwidth]{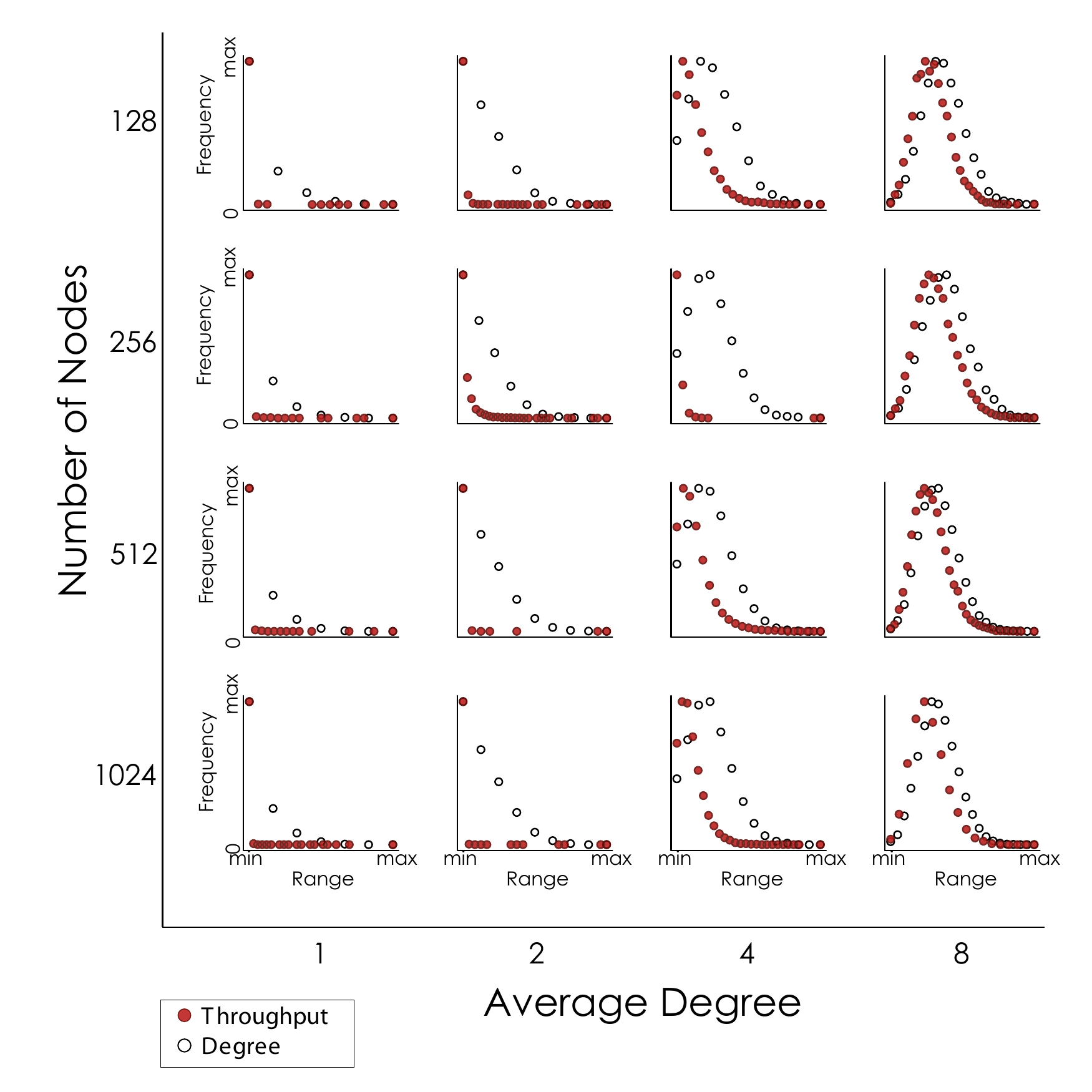}
	\caption{\footnotesize  Throughput is more skewed than degree in sparse networks.  We simulated diffusion of flow in static, directed Erd\"{o}s-R\'{e}nyi random graphs and plotted the distribution of throughput over the nodes (red) in comparison with the distribution of outdegree (white).  Distributions represent totals for fifty networks with each combination of values for average degree and number of nodes, and are normalized over range and frequency.}
	\label{fig:TvD}
\end{figure}

\subsection{The distributions of flow and degree in dynamic random flow networks tend to a power law}  Descriptive statistics of the simulations are presented in Tables \ref{tab:FractionOfTimeInPLFDs} \& \ref{PLDDtab}. Most (98\% overall) simulation runs achieved a power law distribution of flow, and a subset of those reached the PLDD (32\% overall). The highest incidence of power law distributions of flow and degree was for $k_{initial}=4$, while the greatest fraction of time spent in the power law was for $k_{initial}=2$. The apparent reasons for not reaching the power law distribution varied with $k_{initial}$.  With a low initial average degree ($k_{initial}=2$), many networks collapsed quickly, before being observed to reach a PLDD.  With high initial average degree ($k_{initial}=8$), many networks ran to the limit of 50,000 iterations without achieving the PLDD. A possible explanation is that in the initial Erd\"{o}s-R\'{e}nyi random graph, the initial distribution of flow was not skewed enough to give any particular nodes a positional advantage, as indicated in Figure 2.  In all combinations of initial parameters, fewer runs reached a PLDD than a power law flow distribution, consistent with the posited mechanism of the flow distribution limiting the degree distribution.  Quite notably, although most runs achieved power law distributions of flow and many achieved the PLDD, the total time spent in the power law distribution was vanishingly small.  This is considered separately, below.

   \begin{table}[h]
		\centering		
\caption{Characterization of emergent power law flow distributions }
		\begin{tabular}{c@{\vrule height 3pt depth 1pt width0pt}  c c c c c }
		\\		
				\hline
		
		&\multicolumn{4}{c}{$n_{total}$}&\\
				$k_{initial}$&128&256&512&1024&	\\
				\cline{2-5} 
	&\multicolumn{4}{c}{\small{Fraction of runs achieving P.L.}}&\\
						2&1.00&1.00&0.95&0.90\\
			4&1.00&1.00&1.00&1.00\\
			8&1.00&1.00&1.00&0.90\\
		\\
	&\multicolumn{4}{c}{\small{Mean fraction of time in P.L.}}&\\%
			2&0.044&0.020&0.007&0.003\\
				4&0.025&0.007&0.002&$<$0.001\\
				8&0.028&0.009&0.002&$<$0.001\\
\\
			\hline
		\end{tabular}
			\label{tab:FractionOfTimeInPLFDs}
\end{table}

\begin{table}[h]
		\centering		
		\caption{Characterization of emergent PLDDs }
	
		\begin{tabular}{c@{\vrule height 3pt depth 1pt width0pt} c c c c c}

	\\
				\hline
		&\multicolumn{4}{c}{$n_{total}$}&\\
		
				$k_{initial}$&128&256&512&1024&	\\
				\cline{2-5} 
				
				&\multicolumn{4}{c}{\small{Fraction of runs achieving PLDD}}&\\
				2& 0.21& 0.21& 0.12& 0.15\\
				4&0.92 &0.71& 0.58& 0.38\\
				8&0.48& 0.03& 0.00& 0.00\\
				\\	7
		&\multicolumn{4}{c}{\small{Mean fraction of time in PLDD}}&\\
			2&0.216&0.171&0.130&0.026\\
			4&0.035&0.014&0.005&0.002\\
			8&0.006&0.006&-&-\\
		\\
				&	\multicolumn{4}{c}{\small{Mean degree at onset of PLDD}}&\\
			2&2.953&3.248&3.365&3.430\\
			4&3.107&3.364&3.349&3.477\\
			8&3.588&4.369&-&-\\
\\			
			\hline
		\end{tabular}
	\label{PLDDtab}
\end{table}      

\begin{figure}[h!]
	\centering
		\includegraphics[width=0.9\textwidth]{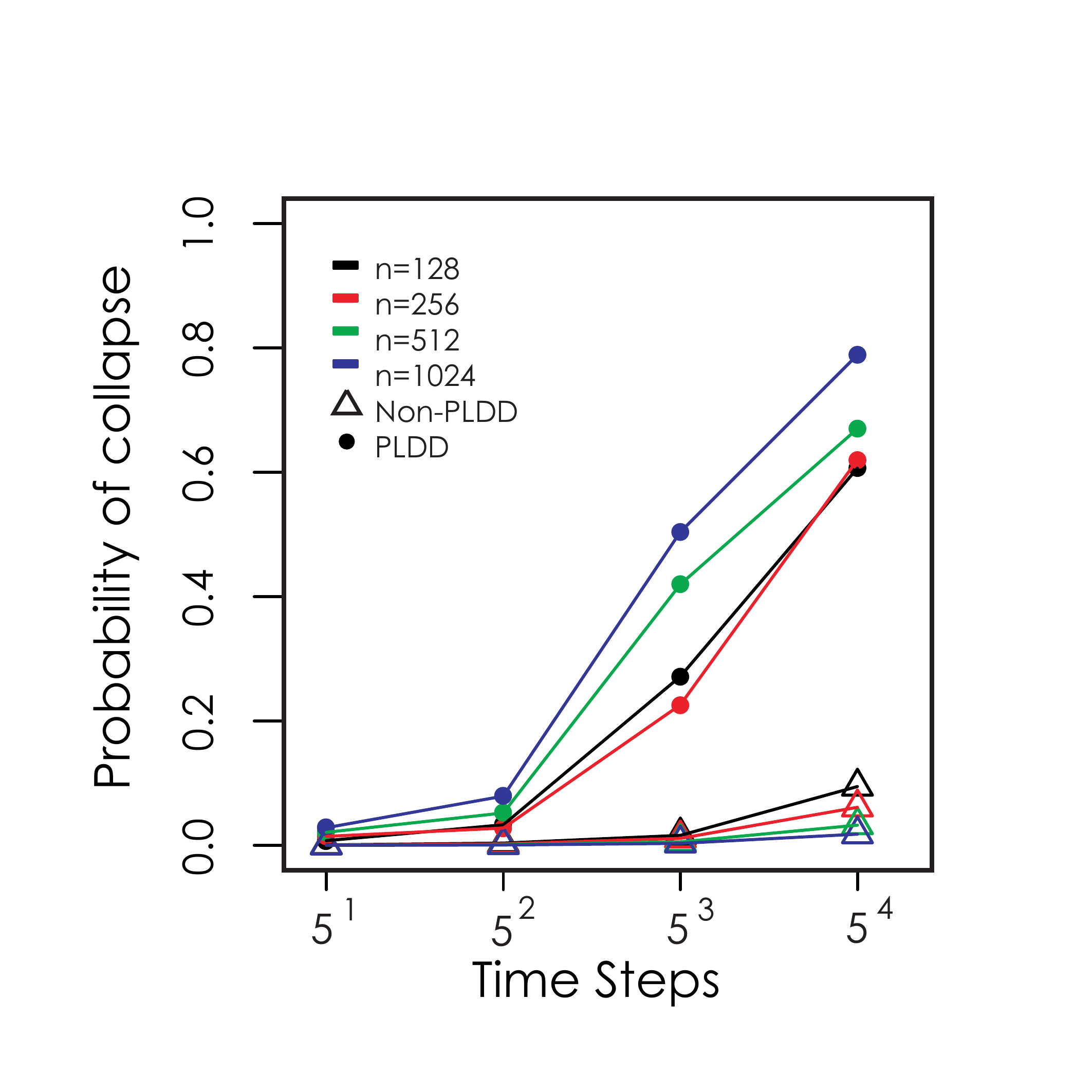}
			\caption{\footnotesize  Characterization of how likely the network is to have collapsed within the number of time steps on the $x-$axis.  Networks in PLDDs (especially with larger $n$) are much more fragile than networks not in PLDD.}
	\label{probcollapse}
\end{figure}

\paragraph{Interpretation} High-flow nodes must be places where the flow from a larger than average ``catchment area'' of upstream nodes becomes concentrated.  A catchment area can be large in the number of nodes or the share of those nodes' throughput that is directed downstream toward a given node.  High flow regions therefore have an above-average chance of being downstream of new links, which tends to maintain or reinforce the skewness of the overall distribution of flow.  Therefore, if new links continue to be formed (assumption 2), even following a uniformly random process, the result is a tendency toward greater concentration of flow in certain regions of the network.  Analogous to self-reinforcing processes in degree producing a PLDD in earlier literature, this self reinforcing process in flow and position produces a power law distribution of flow over the nodes.

If relationships carrying only a negligible flow cannot be maintained, then not only out\textit{flow}, but also out\textit{degree} is limited by inflow. That is, if there is a minimum feasible flow for a given node at a given time (assumption 3), then a node's throughput must establish an upper bound on its outdegree.  For a given threshold, $\iota$, and a level of flow, $f_{i}$, a node's maximum possible outdegree is simply $f_{i}/\iota$.  For very low absolute flow relationships, a further reduction of flow could mean the level of flow passes below that threshold, resulting in the link being pruned.  Removal of a link would also have a broader influence on the network, because it would decrement the throughput of nodes downstream of the dissolution.  Thus, the outdegree-limiting effects of link pruning can cascade through the network, with potential consequences on the global network structure \citep{may2008complex,Beyeler2007693,buldyrev2010catastrophic,albert2004structural}.  Given that the degree of individual nodes gradually increases due to random link creation, the distribution of flow limits the distribution of outdegree, and the prediction that the distribution of flow will tend to a power law, the distribution of outdegrees should therefore also tend in the direction of a power law.

\subsection{Flow networks in a PLDD are at high risk of collapse due to cascading link failure}  The total time spent in the power law was very low, due to the connection between the PLDD and a collapse of a large number of links.  When the networks were in a PLDD, they were much more prone than non-PLDD networks to a catastrophic cascade of links being removed until there were no more active nodes remaining.  For an additional set of 400 runs with $k_{initial}=4$, we took the vector of KS test results (whether the network was in PLDD, tested every 5 iterations), and recorded the length of time from each test result until network collapse, if any.  Figure 3 summarizes these results for moments when the network was in a PLDD separately from when it was not. PLDD networks are dramatically shorter lived than non-PLDD networks, consistent with the hypothesis that they are more prone to collapse.  

\paragraph{Interpretation} According to the interpretation of the previous finding, the PLDD would only occur when the degree distribution has been limited by the flow distribution. This further implies that there are a large number of links with  a flow only slightly above  the minimum flow.  If there are many links with only marginally supracritical flow, we would predict that any random failures would have a higher probability of propagating downstream, because reductions of inflow to a given node are more likely to result in sub-critical flow over its outgoing links.   Therefore, the network is prone to cascading failures when in the PLDD.

\subsection{The algorithm does not produce a rich-get-richer process in degree}
\begin{figure}[h!]
	\centering
		\includegraphics[width=1.0\textwidth]{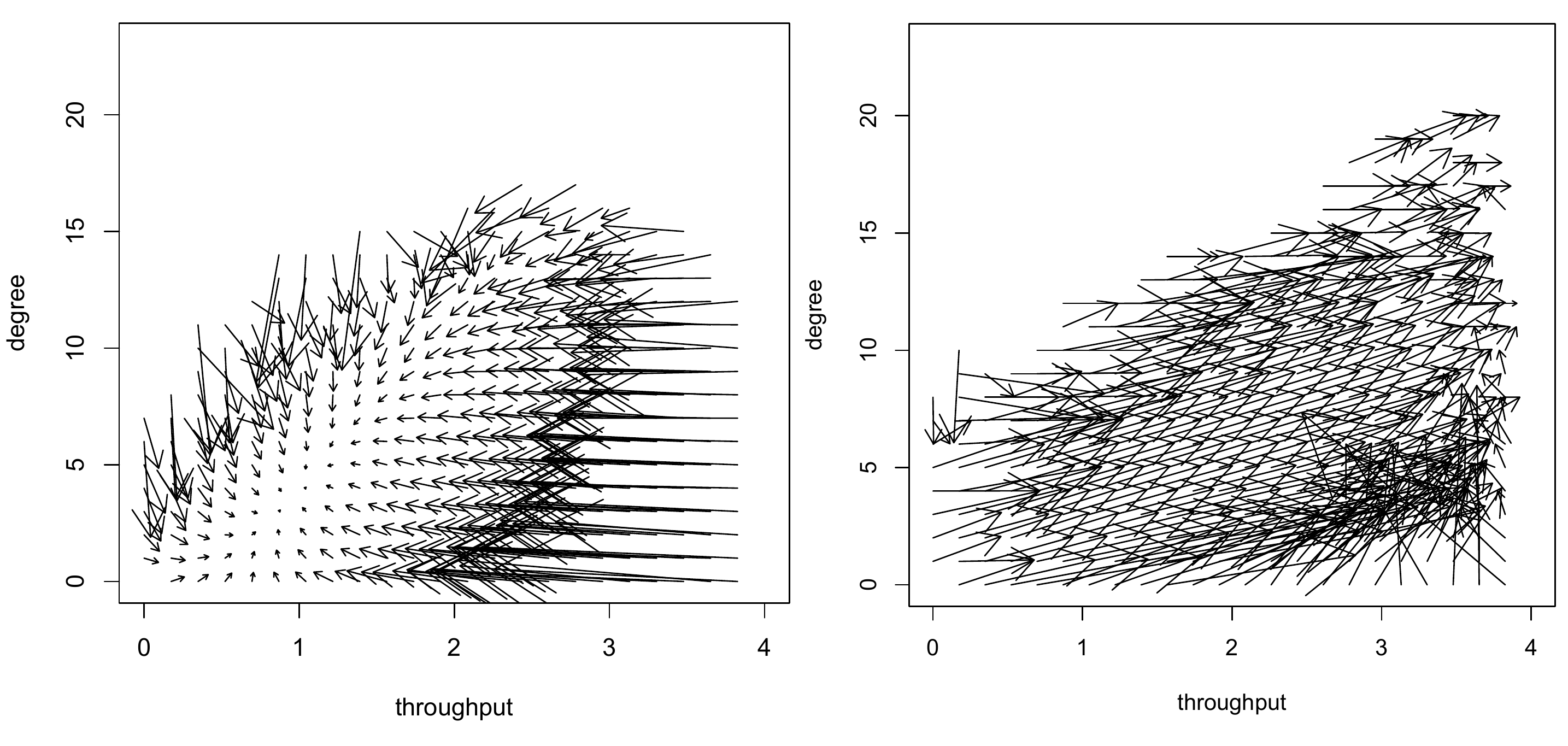}
	\caption{\footnotesize  Average changes (Left) and top 5th percentile increases (Right) in throughput and degree. The arrows show average positions fifty iterations later, for nodes with given values of throughput and degree.  The left panel shows the average of all movements in the sample.  The right panel shows the average movements of the nodes above the 95th percentile increases of throughput and degree for their current values.  These plots only include iterations with 512 total links in the network or fewer.}
	\label{fig:ExpectedChanges}
\end{figure}

Figure \ref{fig:ExpectedChanges} shows average changes in throughput and degree (left panel), and the the top 5 percentile changes in throughput and degree (right panel) on the nodal level for networks with 128 nodes. Because our findings for propositions one through four show that the number of links appears to have a major effect, these figures only represent the subset of iterations where the total number of links in the network was less than 512 (average degree = 4), where the PLDD was most evident for 128 node networks.  

The average behavior is to gravitate to low throughput, low degree.  The higher the node's degree, the faster its degree is expected to decline (that is, we find that our algorithm is actually rich-get-poorer on average).  What we found was that only a small minority of nodes at each level of throughput and degree dependably increases on both of those scales.  

\paragraph{Interpretation} In rich-get-richer algorithms, the higher the current degree, the greater the expected increase in degree. In our model, the addition of another outgoing link does not have a clear relationship with the expected future change in degree.  On one hand, one might see a high current degree as evidence of a relatively high throughput, and thus predictive of an increase in degree correlated with the hypothesized increase in throughput.  On the other hand, adding an outgoing link would divide current throughput among a greater number of links, lowering the flow of each one.  Therefore, the immediate effect of an increase in outdegree could be a greater probability of a loss of one of those links if it becomes too low-flow to sustain. Although high degree is correlated with high throughput, there is a causal counterbalancing force, and the algorithm is not rich get richer in degree overall. 

In summary, the selective pruning of low-flow links has a tendency to result in an increasingly skewed outdegree distribution, and a corresponding increasing susceptibility to catastrophic collapse.  This process frequently skews the outdegree distribution as far as the power law distribution, in which state the network is extremely fragile.

\section{Discussion}

We have shown how minimal assumptions about the nature of flow networks result in a tendency toward skewed, frequently power law, degree distributions. In sparse flow networks, inequality or skewness of the distribution of connections or resources may not be a product of local behavior of individual nodes, but rather be primarily attributable to global structural forces.  That is, nodes with the same local structure can have dramatically different levels of throughput, depending on the greater context in which they are embedded.  The implication for social networks is that an inequitable division of resources, typified by the power law distribution \citep{pareto1964cours}, is actually a probable result of the mechanics of flow, \textit{even without} the constructive aspects of structural dynamics documented in previous network literature that cause rich nodes to get richer \citep{BarAlb1999,bianconi2001competition,kleinberg1999web,saramaki2004scale,vazquez2003growing}. Quantitative sociological models of groups and populations \citep{blau1977macrosociological,mayhew1976emergence} have made similar arguments, but we are not aware of quantitative models within the network paradigm that attribute skewed degree or resource distributions to \textit{global} forces (embedded position with the greater pattern of flow), rather than emergent from aggregated \textit{local} structure and behavior (high degree predisposing nodes to even higher degree). 

Our second major finding was that the PLDD is exceptionally fragile.  Although prior work on the robustness of PLDDs has shown that they are very sensitive to \emph{targeted }attack on or failure of the highest degree hubs \citep{albert2000error,albert2004structural}, we found that in the context of flow networks, \emph{any} attack is likely to affect the network's hubs, because they are necessarily places where a network's flow is concentrated.  Moreover, hubs are likely to rapidly distribute the consequences of failures via their many outgoing links.  In our simulations, we removed the \textit{weakest} links in each iteration and still found networks with a PLDD to be much more fragile than networks without a PLDD (Figure 3).  Similarly, an important instigator of the 2008 global financial crisis was the failure of many small and marginal financial links (sub-prime mortgages), which was then spread throughout the rest of the financial system through several highly connected financial intermediaries \citep{reinhart20082007}.  

Our results do not show whether the instability is a feature of all highly skewed distributions, or only the PLDD.  However, according to our reasoning, vulnerability occurs when the inflow of nodes is just enough to support their outdegrees.  In addition, generally speaking, the more nodes for which outdegree is limited by flow, the more skewed the overall degree distribution will be.  This would further imply that in flow networks, the more skewed the degree distribution, the more vulnerable to collapse the network would be.  

In the context of PLDD-generating algorithms, the distinguishing feature of this model is that it does not rely on rich-get-richer degree dynamics, but rather on a self-reinforcing positional advantage in throughput.  Although the degree distribution is limited by the distribution of flow, this dynamic is distinct from the rich-get-richer processes in prior literature, where change in degree is a probabilistic function of degree itself.  Moreover, in flow networks, rich-get-richer dynamics for outdegree are generally implausible, as change in outdegree depends on in\textit{flow} and is essentially causally independent from current out\textit{degree}. Indeed, we find that on the average, high degree nodes tend to decrease their degree, perhaps because the immediate local effect of an increase in a node's outdegree is a greater chance of future decrease in outdegree, because a new outgoing link weakens a node's other outgoing links by taking a fraction of their flow.  Additionally, we found that relative sparseness was a prerequisite for the tendency toward the PLDD, which was an unexpected point of agreement with a broad study of food webs that found that only sparsely connected webs had a PLDD, while denser ones showed decreasingly skewed functional forms \citep{Dunne01102002}.

Among the limitations of the present study, our model of flow presupposes that the rate of link change is much slower than the diffusion of flow.  Future work should consider networks in which flow does not reach a steady state between link changes.  Additionally, in order to demonstrate generality, we have considered networks with different combinations of initial parameter values over a relatively large number of iterations.  Future work should also examine the scaling properties of flow dynamics across orders of magnitude of network sizes. Our model should also be seen as complementing, rather than contradicting the existing literature.  The bulk of existing literature describes the genesis of networks and their emergent topology, based on incremental growth algorithms.  In contrast, we concentrate on the dynamics of existing systems, and focus on embeddedness within a changing pattern of flow.   In real networks, both sets of dynamics could be operating at once. 

By documenting topological tendencies in general models of flow networks, we hope that our results facilitate interpretation of empirical research on specific systems characterized by flow. Future research on specific flow networks must consider the null hypothesis that skewed distributions of flow and degree over the nodes is a probable outcome due to the general nature of flow networks, and not necessarily due to a form of richer-get-richer degree dynamics.

\section*{References Cited}
\bibliographystyle{model2-names} 

\bibliography{PLnFFlowNetworks.bib}

\end{document}